\documentclass[a4paper,fleqn,usenatbib]{mnras}

\usepackage[T1]{fontenc}
\usepackage{ae,aecompl}

\usepackage{graphicx}	
\usepackage{amsmath}	
\usepackage{amssymb}	

\usepackage{color}
\definecolor{verde}{cmyk}{1,0,1,0}

\title[Lyapunov exponents of periodic orbits]
{The Lyapunov exponents and the neighbourhood of periodic orbits}

\author[D. D. Carpintero and J. C. Muzzio]{
D. D. Carpintero,$^{1,2}$\thanks{E-mail: ddc@fcaglp.unlp.edu.ar (DDC)}
and J. C. Muzzio$^{1,2}$
\\
$^{1}$Facultad de Ciencias Astron\'omicas y Geof\'isicas, Universidad Nacional
de La Plata, Paseo del Bosque s/n, 1900 La Plata, Argentina\\
$^{2}$Instituto de Astrof\'isica de La Plata -- UNLP-Conicet, Paseo del Bosque
s/n, 1900 La Plata, Argentina\\
}

\date{Accepted XXX. Received YYY; in original form ZZZ}

\pubyear{2015}

\begin{document}
\label{firstpage}
\pagerange{\pageref{firstpage}--\pageref{lastpage}}
\maketitle

\begin{abstract}
We show that the Lyapunov exponents of a periodic
orbit can be easily obtained from the eigenvalues of the monodromy matrix.
It turns out that the Lyapunov exponents of simply stable
periodic orbits are all zero, simply unstable periodic orbits have only one
positive Lyapunov exponent, doubly unstable periodic orbits have two
different positive Lyapunov exponents and the two positive Lyapunov
exponents of complex unstable periodic orbits are equal. We present a numerical
example for periodic orbits in a realistic galactic potential. Moreover,
the center manifold theorem allowed us to show that stable, simply unstable
and doubly unstable periodic orbits are the mothers of families of, respectively,
regular, partially and fully chaotic orbits in their neighbourhood.
\end{abstract}

\begin{keywords}
chaos -- instabilities -- galaxies: kinematics and dynamics
\end{keywords}



\section{Introduction}

In several previous articles \citep*{MCW05,M06,MNZ09,ZM12,CMN14,CM16} we have
investigated the role that chaos plays in the dynamics of stellar systems. Since
these systems can be described by autonomous Hamiltonians, their orbits have
always two Lyapunov exponents equal to zero, and the remaining four are always
two pairs of opposite real numbers \citep[e.g.,][]{BGS76}. This means that there
may be zero, one or two
positive Lyapunov exponents. No positive Lyapunov exponents means that there is
no direction in phase space along which two initially infinitesimally separated
orbits diverge exponentially, that is, the original orbit is regular. Otherwise,
the orbit is chaotic. But chaotic orbits are evidently not all the same: there
are those with only one positive Lyapunov exponent, called partially chaotic
orbits, and those with two positive Lyapunov exponents,
called fully chaotic orbits. 
One of the main results from our abovementioned works was that fully chaotic
orbits are a disjoint family from the partially chaotic orbits, even in their
spatial distribution inside the system.

\citet*{CGG78} and \citet{PV84} had also reported
the finding of partially chaotic orbits in other autonomous Hamiltonian
systems, but their existence was denied by \citet{F70, F71} and by
\citet{LL92}. Nevertheless, more recently, \citet{M17, M18} proved their
existence over, at least, time intervals of 50 million
Hubble times.

On the other hand, studies of the stability of periodic orbits in the three-body
problem \citep{H75} and in triaxial potentials \citep{M82,CM85,C02} have yielded a wealth of
phenomena of interest. The different classes of instability of these orbits are
determined according to the eigenvalues of the monodromy matrix of the periodic
orbit. In particular, \cite{CM85} have classified the orbits into four
categories: stable, unstable, doubly unstable and complex unstable. The question
naturally arises of whether these orbits are somehow related
to the regular, partially and fully chaotic orbits. Therefore, we decided to
investigate the relationship between the eigenvalues of the monodromy matrix of
periodic orbits and the Lyapunov exponents of those orbits and we found that, in
fact, the latter can be computed from the former. Besides, the
stability of the periodic orbits, influences decisively the phase space in their
neighbourhood and it turns out that, just as stable periodic orbits are surrounded
by regular orbits, simply unstable periodic orbits are surrounded by partially
chaotic orbits and doubly and complex unstable periodic orbits are surrounded
by fully chaotic orbits. The present paper presents our results. 
Section \ref{sdos} gives our analytical proof, section \ref{stres}
presents a numerical example and our conclusions are described in Section \ref{scuatro}.

\section{Lyapunov exponents and periodic motion}
\label{sdos}

\subsection{Lyapunov exponents as eigenvalues of the main fundamental matrix}

In a 3D potential, a regular orbit is defined as an orbit which obeys al least
three isolating integrals of motion \citep[e.g.,][]{BT08}; otherwise, it is irregular. On
the other hand, a chaotic orbit is defined as an orbit which has sensitivity to
the initial conditions, that is, if its initial phase space position 
$\mathbfit{w}_0\equiv (\mathbfit{x}_0,\mathbfit{v}_0)$ is infinitesimally
perturbed, then the new  orbit (hereafter called perturbed orbit) diverges
exponentially from the original one. Though there is no proof that every irregular
orbit is chaotic, we will stick to the widespread custom of
considering both sets as the same.

The standard gauge to measure the rate of divergence between an orbit and its
perturbed 
sister is the set of Lyapunov exponents, sometimes called Lyapunov
characteristic numbers or Lyapunov characteristic exponents.
If we choose six independent directions
of the phase space $\mathbfit{e}_i$, $i=1,\dots,6$, and perturb the initial
conditions in the direction $j$ by the amount $\delta w_j(0)$, then
the Lyapunov 
exponents are defined by \citep[e.g.,][]{LL92}
\begin{equation}
\label{uno}
 \lambda_i=\lim_{t\to\infty}{1\over t} \ln{|\delta w_i(t)| \over
|\delta w_i(0)|},\quad i=1,\dots,6, 
\end{equation}
where $\delta w_i(t)$ is the component of the deviation at time $t$ of the initial $i$th component of the
deviation 
$\delta w_i(0)$, and the norm $|\cdot|$ is any norm of the phase
space, 
normally the Euclidean one.

Numerically, the deviation $\delta\mathbfit{w}(t)$ can be computed starting from the equations of motion
\begin{equation}
\label{eqm}
 \mathbfit{\.w} = \mathbfit{F}(\mathbfit{w}),
\end{equation}
where $\mathbfit{w}(t)$ is the phase point of the orbit, $\mathbfit{\.w}(t)$ its velocity, and $\mathbfit{F}$ are the functions that define the dynamical system. Developing Eqs. (\ref{eqm})
in a Taylor series around the unperturbed orbit and retaining only the first
order, we obtain the so called variational 
equations:
\begin{equation}
\label{var}
 {\text{d}\over\text{d}t}(\delta\mathbfit{w})
 =\left. {\upartial\mathbfit{F}\over\upartial\mathbfit{w}}
 \right|_\mathbfit{w}\cdot
   \delta\mathbfit{w}.
\end{equation}

The set of variational equations (\ref{var}) for the six components of the deviation is,
then, a system of 
linear, homogeneous, 
ordinary differential equations. For this kind of systems, a fundamental matrix
$\mathbfss{S}(t)$ is
defined as a matrix whose columns are linearly independent solutions of
the system \citep[e.g.,][]{RS76}. In our case,
\begin{equation}
 \mathbfss{S}(t)=(\delta\mathbfit{w}_1,\dots,\delta\mathbfit{w}_6),
\end{equation}
where each $\delta\mathbfit{w}_i$ is an independent solution of Eq. (\ref{var}) and represents a column,
is the fundamental matrix of the variational equations. It is then clear that
\begin{equation}
\label{ocho}
 \mathbfss{\.S}(t)
 =\left. {\upartial\mathbfit{F}\over\upartial\mathbfit{w}}
 \right|_\mathbfit{w}\cdot
   \mathbfss{S}(t).
\end{equation}
Since the $\delta\mathbfit{w}_i$ are linearly independent solutions, then
\begin{equation}
\mathbfss{S}\cdot \mathbfss{S}^{\textsf T}=
\text{diag}(|\delta\mathbfit{w}_1|^2, \dots,|\delta\mathbfit{w}_6|^2)=\mathbfss{S}^2,
\end{equation}
where $(\cdot)^{\textsf T}$ indicates transposition. 
If we choose $\mathbfss{S}(t)$ such that $\mathbfss{S}(0)=\mathbfss{1}$, i.e.,
the identity
matrix (in which case $\mathbfss{S}(t)$ is called main fundamental matrix), 
Eq. (\ref{uno}) shows that 
the set of Lyapunov exponents can be expressed as the eigenvalues
of a (diagonal) matrix $\mathbfss{L}$ where
\begin{equation}
\label{lyapu}
 \mathbfss{L}=\lim_{t\to\infty}{1\over 2t} 
\ln(\mathbfss{S}^2)=
\lim_{t\to\infty}{1\over t} 
\ln|\mathbfss{S}|,
\end{equation}
where $|\mathbfss{S}|=\text{diag}(|\delta\mathbfit{w}_1|, \dots,|\delta\mathbfit{w}_6|)$ \citep[e.g.,][]{BGGS80a}.

Now, let
\begin{equation}
\mathbfit{w}(t)=\mathbfit{G}(t,\mathbfit{w}_0)
\end{equation}
be the solution of Eq. (\ref{eqm}) with initial condition $\mathbfit{w}(0)=\mathbfit{w}_0$, and
let the matrix $\mathbfss{M}$ be defined by
\begin{equation}
 \mathbfss{M}(t)={\upartial 
 \mathbfit{G}(t,\mathbfit{w}_0)\over
 \upartial \mathbfit{w}_0}=
{\upartial \mathbfit{w}(t)\over
 \upartial \mathbfit{w}_0},
\end{equation}
i.e. $\mathbfss{M}(t)$ is the matrix that evolves the initial perturbation until time $t$:
\begin{equation}
\label{evin}
\delta\mathbfit{w}(t)=\mathbfss{M}(t)
\cdot\delta\mathbfit{w}_0.
\end{equation}
Now, by applying the chain rule, we have
\begin{equation}
\label{eme}
\mathbfss{\.M}=
{\upartial \mathbfit{\.w}\over \upartial \mathbfit{w}_0} =
{\upartial \mathbfit{\.w}\over \upartial \mathbfit{w}} \cdot
{\upartial \mathbfit{w}\over \upartial \mathbfit{w}_0} =
{\upartial \mathbfit{F}\over \upartial \mathbfit{w}} \cdot
\mathbfss{M},
\end{equation}
i.e., $\mathbfss{M}$ turns out to be the fundamental matrix $\mathbfss{S}$ of the variational equations (cf. Eq. (\ref{ocho})).

\subsection{Lyapunov exponents of a periodic motion}

We now specialise in periodic motion. Let the solution of Eq. (\ref{eqm}) represent a periodic orbit of period $T$. The stability of such an
orbit can be established studying the behavior of a second orbit
obtained by perturbing the initial conditions, i.e. by integrating the variational equations. Let $\mathbfss{M}$ be the fundamental matrix of this system; it satisfies (cf. Eq. (\ref{eme}))
\begin{equation}
 \mathbfss{\.M}(t)
 =\left. {\upartial\mathbfit{\.w}\over\upartial\mathbfit{w}}
 \right|_\mathbfit{w}\cdot
   \mathbfss{M}(t).
\end{equation}

According to the
Floquet theorem \citep{F883}, the fundamental matrix $\mathbfss{M}(t)$ in this case is also
periodic with period $T$. This property, along with Eq.
(\ref{evin}) and the assumption that
$\mathbfss{M}(0)=\mathbfss{1}$ allow us to write
\begin{equation}
 \delta\mathbfit{w}(T)=\mathbfss{M}(T)\cdot\delta\mathbfit{w}(0).
\end{equation}
The main fundamental matrix evaluated at $t=T$, $\mathbfss{M}(T)$, is called the
monodromy matrix of the periodic orbit \citep[e.g.][]{C02}. 

Since the motion is periodic with period $T$, then
\begin{equation}
\begin{split}
 \delta\mathbfit{w}(2T) & = \mathbfss{M}(2T)\cdot\delta\mathbfit{w}(0) \\
 & =\mathbfss{M}(T)\cdot\left[\mathbfss{M}(T)\cdot\delta\mathbfit{w}(0)\right] \\
 & =\left[\mathbfss{M}(T)\right]^2\cdot\delta\mathbfit{w}(0),
\end{split}
\end{equation}
so we have, for $n\in{\mathbb N}$,
\begin{equation}
\label{pote}
 \mathbfss{M}(nT)=\left[\mathbfss{M}(T)\right]^n.
\end{equation}
Then, the Lyapunov exponents can be easily computed as the natural logarithms of the eigenvalues of the monodromy matrix divided by $T$ (cf. Eq.(\ref{lyapu})):
\begin{equation}
\label{lymo}
\begin{split}
 \mathbfss{L}&=\lim_{t\to\infty}{1\over t} 
\ln\left|\mathbfss{M}(t)\right| \\
&=\lim_{n\to\infty}{1\over nT} 
\ln\left|\mathbfss{M}(nT)\right| \\
&={1\over T}\ln\left|\mathbfss{M}(T)\right|,
\end{split}
\end{equation}
where in the last line we have used Eq. (\ref{pote}). 
If we let
$\ell_i$, $i=1,\dots,6$ be the six eigenvalues of $\mathbfss{M}(T)$, then we have
\begin{equation}
\label{lymoli}
\lambda_i={1\over T}\ln|\ell_i|.
\end{equation}

As usual, the $\ell_i$'s are obtained as the six roots of the characteristic polynomial of  $\mathbfss{M}(T)$,
\begin{equation}
 a_6\ell^6 + a_5\ell^5 + a_4\ell^4 + a_3\ell^3 + a_2\ell^2 + a_1\ell + a_0=0,
\end{equation}
where the $a_i$ are real numbers. The two null Lyapunov exponents imply that two of the eigenvalues are always unity. After dividing by $(\ell-1)^2$, the remaining polynomial we write as
\begin{equation}
\label{poli}
 c_4\ell^4 + c_3\ell^3 + c_2\ell^2 + c_1\ell + c_0=0.
\end{equation}
We also know that the four remaining Lyapunov exponents come in pairs of opposite numbers, so the corresponding eigenvalues are pairs of reciprocal numbers. Let $\{\ell_1,\ell_2=\ell_1^{-1},\ell_3,\ell_4=\ell_3^{-1}\}$ be those eigenvalues. Then, the polynomial (\ref{poli}) can be written
\begin{equation}
\label{cuarto}
    \ell^4+\alpha\ell^3+\beta\ell^2+\alpha\ell+1=0,
\end{equation}
where
\begin{equation}
\begin{split}
    \alpha&=-(\ell_1+\ell_2+\ell_3+\ell_4), \\
    \beta&=\ell_1\ell_3+\ell_1\ell_4+\ell_2\ell_3+
    \ell_2\ell_4+2.
\end{split}
\end{equation}
Following the standard notation initiated by \cite{H75}, we let
\begin{equation}
\label{bes}
\begin{split}
b_1&=-(\ell_1+\ell_2), \\
b_2&=-(\ell_3+\ell_4),
\end{split}
\end{equation}
with which
\begin{equation}
\begin{split}
\alpha&=b_1 + b_2, \\
\beta&=b_1 b_2+2,
\end{split}
\end{equation}
and, therefore,
\begin{equation}
\begin{split}
b_1 & = {1\over 2}
    \left(\alpha + \sqrt{\Delta}\right), \\
 b_2& = {1\over 2}
    \left(\alpha - \sqrt{\Delta}\right),
\end{split}
\end{equation}
where
\begin{equation}
\Delta  = \alpha^2-4(\beta-2).
\end{equation}
With this notation, the four roots (eigenvalues) can be written as
\begin{equation}
\begin{split}
    \ell_{1,2}&={1\over 2}     \left(-b_1\pm\sqrt{b_1^2-4}\right),\\
    \ell_{3,4}&={1\over 2}     \left(-b_2\pm\sqrt{b_2^2-4}\right).
\end{split}
\end{equation}

\subsection{Stability of a periodic motion}

Although the
Lyapunov exponents of a periodic motion can be effortlessly computed from the
eigenvalues of the monodromy matrix, the stability of its
orbits is usually studied by considering $\Delta$, $b_1$ and $b_2$ \citep[e.g.][]{H75,M82,CM85,PZ90,PZ94,C02}, which are combinations of those eigenvalues:
\begin{equation}
\Delta=(-\ell_1-\ell_2+\ell_3+\ell_4)^2
\end{equation}
and $b_1$, $b_2$ given by Eq. (\ref{bes}). Using this three indicators, a periodic orbit is found to be \citep[e.g.,][]{C02}:
\begin{enumerate}
\item \emph{Stable}, if $\Delta>0$, $|b_1|<2$, and $|b_2|<2$. In this case, all the eigenvalues are complex numbers lying on the unit circle, and, besides their reciprocal property $\ell_1\ell_2=\ell_3\ell_4=1$, the pairs also obey
$\ell_1=\ell_2^*$ and $\ell_3=\ell_4^*$,
where the asterisk means complex conjugation.
\item \emph{Unstable}, if $\Delta>0$, $|b_1|<2$, and $|b_2|>2$ or $\Delta>0$, $|b_1|>2$, and $|b_2|<2$. In this case, a pair of reciprocal roots are complex conjugate lying on the unit circle, and the other two roots are real.
\item \emph{Doubly Unstable}, if $\Delta>0$, $|b_1|>2$, and $|b_2|>2$. All four roots are real.
\item \emph{Complex Unstable}, if $\Delta<0$. In this case, besides $\ell_1\ell_2=\ell_3\ell_4=1$, we have $\ell_1=\ell_3^*$ and $\ell_2=\ell_4^*$,
that is, the reciprocal and conjugate pairs are different.
\end{enumerate}

Now, we want to write these four types of stability in terms of the Lyapunov exponents. According to Eq. (\ref{lymoli}), any pair of reciprocal eigenvalues $\ell_i=\ell_j^{-1}$ will yield
\begin{equation}
\lambda_i={1\over T}\ln|\ell_i|=
{1\over T}\ln|\ell_j^{-1}|=-{1\over T}\ln|\ell_j|=-\lambda_j,
\end{equation}
and any pair of conjugate eigenvalues $\ell_i=\ell_j^*$ will yield
\begin{equation}
\lambda_i={1\over T}\ln|\ell_i|=
{1\over T}\ln|\ell_j^*|={1\over T}\ln|\ell_j|=\lambda_j.
\end{equation}

The different stability cases then yield:
\begin{enumerate}

\item Stable orbits. Both pairs of roots are simultaneously reciprocal and conjugate. For the first pair $\{\ell_1,\ell_2\}$ we have
$\lambda_1=-\lambda_2$ and $\lambda_1=\lambda_2$,
and the same for $\{\ell_3,\ell_4\}$. Therefore, 
\begin{equation}
\lambda_1=\lambda_2=\lambda_3=\lambda_4=0,
\end{equation}
that is, stable periodic orbits have all their Lyapunov
exponents equal to zero.

\item Unstable orbits. Two of the roots are reciprocal and  conjugate, so the previous analysis apply. The other two are only reciprocal. Therefore, we have
\begin{equation}
\begin{split}
\lambda_1&=\lambda_2=0,\\
\lambda_3&=-\lambda_4
\end{split}
\end{equation}
if $|b_1|<2$, or with the pairs interchanged if $|b_2|<2$.
Thus, unstable periodic orbits have only two non-zero and
opposite Lyapunov exponents.

\item Doubly Unstable orbits. Now both pairs are only reciprocal, so we have
\begin{equation}
\begin{split}
\lambda_1&=-\lambda_2,\\
\lambda_3&=-\lambda_4.
\end{split}
\end{equation}
Therefore, there are four non-zero Lyapunov exponents that
are opposite in pairs for doubly unstable orbits.

\item Complex Unstable orbits. In this case we have 
$\lambda_1=-\lambda_2$, $\lambda_3=-\lambda_4$, 
$\lambda_1=\lambda_3$, $\lambda_2=\lambda_4$,
and therefore
\begin{equation}
\lambda_1=-\lambda_2=\lambda_3=-\lambda_4,
\end{equation}
that is, a complex unstable orbit has two equal pairs of
opposite non-zero Lyapunov exponents.
\end{enumerate}

\subsection{The neigbourhood of periodic orbits.}

Thus far, we have only dealt with the periodic orbits themselves,
but the theorem of existence of center manifolds \citep*[e.g.][]{GH13,B01}
allows us to extend our results to the neighbourhood of those orbits, and
will lead us to an important conclusion.
Let us consider the 4D Poincar\'e map around the fixed point of
a simply unstable periodic orbit which has one positive, one negative and two
zero eigenvalues. Thus, according to the theorem, there exist in the neighbourhood
of the fixed point a 1D unstable local invariant manifold, a 1D stable local invariant
manifold and a 2D centre invariant manifold\footnote{Figures 1.1.5 and 1.1.6 of
\citet{W90} provide nice 3D examples of orbits for
the case of stable and unstable manifolds.}. Therefore, the Lyapunov
exponents of the orbits in that neighbourhood will be one negative, one positive
and four zero (two due to the centre invariant manifold and two due to the
conservation of energy), i.e., they are partially chaotic orbits. The same
reasoning can be applied to the stable and to the doubly unstable orbits. In brief,
stable and simply and unstable periodic orbits are the mothers of families of,
respectively, regular and partially chaotic orbits, while both doubly and complex unstable periodic
orbits are the mothers of families of fully chaotic orbits.

\section{Numerical example}
\label{stres}

\citet{PZ90} investigated the stability of the
periodic orbits along the axis of rotation of a model galaxy
using the monodromy matrix. They modeled the potential with a
disc of the \citet{MN75} type and a triaxial
logarithmic halo. We used their potential together with the
{\sc Liamag} routine, kindly provided by D. Pfenniger \citep[see][]{UP88}
to obtain the Lyapunov exponents. We selected
initial conditions for orbits along the axis of rotation with
different energies and values of the angular
velocity; the integration time was $10^{6}$ time units. Two
positive Lyapunov exponents were considered equal when they differed by
less than $1.25\times 10^{-4}$.

\begin{figure}
	\includegraphics[width=\columnwidth]{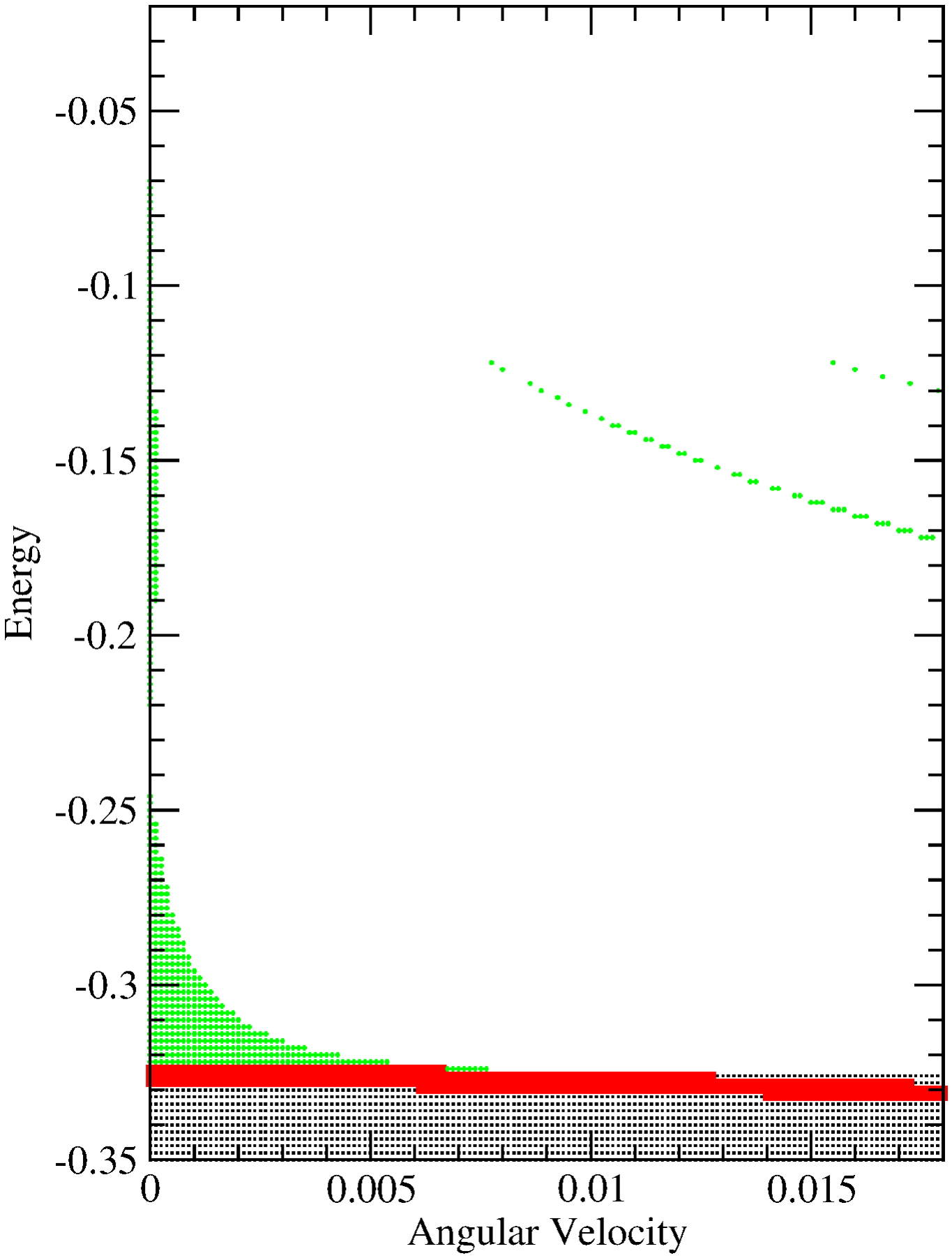}
    \caption{Types of orbits in the angular velocity vs. energy
diagram. Orbits with null Lyapunov exponents are shown
as dots (black in the electronic version) and orbits with only one
positive Lyapunov exponent as filled squares (red in the electronic version).
The blank space
corresponds to orbits with two equal non-zero Lyapunov exponents
and plus signs (green in the electronic version) represent
	orbits with two non-zero Lyapunov exponents that are not equal.}
    \label{fig01}
\end{figure}

Fig.~\ref{fig01} presents our results and it can be compared with
Figure 2 of \citet{PZ90}. The stable and simply unstable regions of
their diagram agree very well with the regions of our
Fig.~\ref{fig01} that correspond, respectively, to our regions with all
null and with just one positive Lyapunov exponents. The comparison of the
doubly unstable and complex unstable regions of Patsis and Zachilas
with our regions occupied by orbits with two positive Lyapunov exponents and, respectively,
$\lambda_{1}\neq\lambda_{2}$ and $\lambda_{1}=\lambda_{2}$ shows, however,
some small disagreements. For a rotationless galaxy our results give
orbits with $\lambda_{1}=\lambda_{2}>0$ for energies between
$-0.245$ and $-0.220$ as well as for energies larger than $-0.070$, although
the results of Patsis and Zachilas give doubly unstable orbits for all
energies. Besides, the two lanes of orbits with two
positive Lyapunov exponents and $\lambda_{1}\neq\lambda_{2}$
in the upper right region of our figure do not extend to energies larger
than $-0.120$, while the corresponding lanes of doubly unstable orbits of
Patsis and Zachilas continue up to the top of their figure. The problem
is that the differences between $\lambda_{1}$ and $\lambda_{2}$ are very
small in those regions and close to the precision of our computations.
The differences between the two positive Lyapunov exponents of the orbits
in the uppermost right lane, for example, are about $1.50\times 10^{-4}$, i.e.
very near our limiting value of $1.25\times 10^{-4}$. The method of the monodromy
matrix seems, therefore, to be better than Lyapunov exponents to distinguish
doubly unstable from complex unstable periodic orbits, but that is not a
problem for us because we did not intend to replace the
method of the monodromy matrix by the use of Lyapunov exponents.

\section{Conclusions}
\label{scuatro}

We have proven that stable periodic orbits have null
Lyapunov exponents, simply unstable periodic orbits have
only one positive Lyapunov exponent, doubly unstable periodic orbits
have two different positive Lyapunov exponents and complex unstable periodic
orbits have two equal positive Lyapunov exponents.
A corollary of our result is that complex instability does not exist in
systems with two-dimensional (2D) configuration spaces, an assertion that
\citet[][p. 287]{C02} gives without proof, because complex instability
demands an exponential expansion in two dimensions but only one is
available in 2D autonomous systems (the other one has a zero Lyapunov
exponent).

The most important result of our study is that the stability
of the periodic orbits (revealed by their Lyapunov exponents) drastically
affects the phase space in their neighborhood and, as a result, stable, simply
unstable and both doubly and complex unstable periodic orbits should be
surrounded by families of, respectively, regular, partially and fully
chaotic orbits. Thus, our result gives further support to the existence
of partially chaotic orbits.

We have investigated the presence of chaos in many galactic
models in the past and our experience is that $\lambda_{1}$ is usually much
larger than $\lambda_{2}$ when both are not zero. Thus, it was surprising to
find that almost 98 per cent of the orbits with two non-zero Lyapunov exponents
in Figure~\ref{fig01} have $\lambda_{1}=\lambda_{2}$.
The most likely explanation for this oddity is that all the orbits in that
sample are periodic, while in our models it would have been almost impossible
to find a periodic orbit by chance.

\section*{Acknowledgements}
The comments of an anonymous referee were very
useful to improve the original version of this paper and are
gratefully acknowledged.
We are very grateful to D. Pfenniger for the use of his code.
We acknowledge support from grants from the Universidad Nacional
de La Plata, Proyecto 11/G153, and from the CONICET, PIP 0426.




\bibliographystyle{mnras}
\bibliography{biblio} 






\bsp	
\label{lastpage}
\end{document}